\documentstyle[12pt,aaspp4]{article}
\lefthead{Aparicio et al.}
\righthead{The SFH of LGS3}
\begin{document}

\title{The stellar content and the Star Formation History of the Local Group Dwarf Galaxy LGS~3
\altaffilmark{1}}

\author{A. Aparicio\altaffilmark{2}}

\affil{Instituto de Astrof\'\i sica de Canarias, E-38200  La Laguna, Tenerife, Canary Islands, Spain}

\author{C. Gallart}

\affil{Observatories of the Carnegie Institution of Washington, 813 Santa Barbara St., Pasadena, CA 91101, USA}

\and

\author{G. Bertelli}
\affil{Dipartimento di Astronomia, Vicolo dell'Osservatorio 5, I35122 Padova, Italy}
\affil{Fellow of the Consiglio delle Ricerche (CNR-GNA), Roma, Italy}

\altaffiltext{1} {Based on observations made with the Nordic Optical Telescope
operated on the island of La Palma in the Spanish Observatorio del Roque de los Muchachos of the Instituto de Astrof\'\i sic
a de Canarias.}

\altaffiltext{2}{Presently at Observatories of the Carnegie Institution of Washington, 813 Santa Barbara St., Pasadena, CA 91101, USA} 

\begin{abstract}

The star formation history (SFH) and the properties of the dwarf galaxy LGS~3 
are analyzed using color-magnitude (CM) diagrams plotted from $VRI$ photometry of 736 
stars. The distance to the galaxy is estimated through the position of the tip 
or the red giant branch. Two acceptable results have been obtained: $0.77\pm 
0.07$ Mpc and $0.96\pm 0.07$ Mpc, although the first value is favored by 
complementary considerations on the stellar content of the galaxy. Both values 
make LGS~3 a possible satellite of M31 or of M33.  
The SFH is investigated for each of the two adopted distances comparing the 
observed (CM) diagrams with model CM diagrams computed for different star formation rates ($\psi(t)$) and chemical enrichment laws ($Z(t)$). 

The results are compatible with LGS~3 having been forming stars since an early epoch, 
15--12 Gyr ago, at an almost constant rate if distance is 0.77 Mpc or at an exponentially decreasing rate if distance is 0.96 Mpc. According to our models, the current metallicity would range from $Z\simeq 0.0007$ to $Z\simeq 0.002$. Other 
results are the current $\psi(t)$: $(0.55\pm 0.04)\times 10^{-10}$ 
M$_\odot$ yr$^{-1}$ pc$^{-2}$ or $(0.47\pm 0.07)\times 10^{-10}$ 
M$_\odot$ yr$^{-1}$ pc$^{-2}$, depending of the distance, and its average for the 
entire life of the galaxy, $\bar\psi=(1.4\pm 0.1)\times 10^{-10}$ 
M$_\odot$ yr$^{-1}$ pc$^{-2}$. At the present $\psi(t)$, the probability of LGS~3 
having an HII region is 0.2, which is compatible with the fact that no HII 
regions have been found in the galaxy. Its fraction of gas relative to the mass 
intervening in the chemical evolution is about 0.40 and its percentage of dark 
matter (that which cannot be explained as stellar remnants or by extrapolation 
of the used IMF to low masses) is 95\%. The results for $\psi(t)$ and $Z(t)$ for $d=0.77$ Mpc are compatible with a moderate outflow of well mixed material ($\lambda=3$), but large outflow rate ($\lambda=30$) is required to account for the results for $d=0.96$ Mpc. The latter would imply that, if the amount of dark mass associated to the galaxy is constant, the initial dark matter fraction was about 50\%. In both cases, a large fraction of freshly made metals ($\gamma=0.85$ for the case of $d=0.77$ Mpc and $\gamma=0.36$ for $d=0.96$ Mpc) should also escape from the galaxy before mixing with the interstellar medium to make the results compatible with the theoretical yields.

LGS~3 is considered an intermediate type between dIr and dE. However we find 
that it shows characteristics typical of dIrs (the $M_{\rm gas}/L_B$ ratio and the SFH shape), the main difference being that its mass and star formation rate (present and averaged) are one to two orders of magnitude smaller than those of other dIrs. 
This makes the absence of HII regions a simple probabilistic effect. Considering 
this and the fact that LGS~3 can continue to form stars for a further 10 Gyr at a rate equal to that averaged for its past history, we conclude that this galaxy may be considered just a dIr in the tail of the distributions of mass, luminosity and star formation 
rate.

\end{abstract}

\section{Introduction}

Dwarfs seem to be the most abundant kind of galaxies in the Universe. That is at least 
the case for 
the Local Group, where all but four of its members are dwarfs. Although the 
light and the 
dynamics of the Local Group are dominated by the two largest galaxies, Andromeda 
and the Milky 
Way, the spatial distribution, velocities and other physical properties of dwarf 
galaxies in the 
Group and its neighborhood should provide information on its dynamical and 
evolutionary properties. In particular, the interconnections of the Milky Way with 
its satellites may contain clues for the mechanisms in which the galaxy was 
formed (see Majewski~\markcite{majewski}1994). Dwarf galaxies could also play an 
important role in 
cosmological models, in particular, in relation to the issue of whether galaxies 
formed in a cold or in a hot dark matter 
scenario. 

Dwarf galaxies can be divided into two large groups: dwarf ellipticals (dEs; we 
include the dwarf spheroidal galaxies in this group) and dwarf irregulars (dIrs). 
Both kinds of dwarfs cover similar ranges of mass and luminosity, the key 
difference between them being, besides morphology, that the latter have a much 
larger amount of gas and young stars that the former, where HII regions are absent. The $M_{\rm H\ I}/L_B$ ratio, which is of the order of 1 for dIrs and 
close to 0 for dEs, can be used to separate between both groups (Da 
Costa~\markcite{mhlb}1997). dEs and dIrs also show a decoupled spatial 
distribution in the Local Group: dEs are preferentially distributed close to the 
largest galaxies, while dIrs are found normally far away from these systems and 
many of them could be field galaxies rather than Local Group members. But the 
boundaries between dEs and dIrs are not sharp. Focusing on the stellar population, 
many dEs show indication of intermediate-age or even young stars (Smecker-Hane et 
al.~\markcite{shane}1996; Freedman~\markcite{freedman}1994, and references 
therein) while most dIrs show the so-called Baade's sheet (Baade 
\markcite{baade}1963) which would be composed by a large number of 
intermediate-age and old stars\footnote{In this paper, we mean by old stars 
those older than 10 Gyr, by intermediate-age stars those with ages between 1 
and 10 Gyr and by young stars  those younger than 1 Gyr.} if the generality of dIr 
have star formation histories (SFH) like those of NGC~6822 (Gallart et 
al.~\markcite{n68hstb}1996b,\markcite{n68hstc}1996c) and Pegasus  (Aparicio, 
Gallart, \& Bertelli \markcite{peg3}1997). A gradation exists in the properties 
from dEs to dIrs and it is argued that LGS~3 may be a transition object between a 
dIr galaxy and a dE (see below).  

LGS~3 was discovered by Karachentseva \markcite{kar} (1976). Thuan \& Martin 
\markcite{thuanmartin} (1979) measured its HI content obtaining a negative 
velocity $-280\pm 8$ km s$^{-1}$ or, relative to the centroid of the Local Group 
(LG), $-0.41\pm 8$ km s$^{-1}$ and an HI mass of $M_{\rm H\ I}=(2.15\pm 0.36)\times 
10^5$ M$_\odot$, assuming the galaxy being at the same distance as M33 (0.817 
Mpc was used). Tifft \& Cocke \markcite{tifftcocke} (1988) obtained new measures 
of HI and a velocity relative to the Sun of $-286$ km s$^{-1}$. Lo, Sargent, \& 
Young \markcite{lo} (1993) estimated the HI mass of LGS to be $(4.0\pm 
0.5)\times 10^5$ M$_\odot$ Mpc$^{-2}$ and its total mass $(1.8\pm 1)\times 10^7$ 
M$_\odot$, for which they adopted a distance of 0.76 Mpc. Early optical studies 
of the galaxy are those by Schild \markcite{schild} (1980), who obtained a value 
for its total magnitude and by Christian \& Tully \markcite{chris} (1983), who 
performed photometry of individual stars and obtained a first independent 
estimate of the distance between 0.7 and 1.2 Mpc. Late studies of LGS~3 are 
those by Lee \markcite{lee} (1995),  Tikhonov \& Makarova \markcite{} (1996), Mould~\markcite{mould}(1997) and Karachentseva et al.~\markcite{kar2}(1997). Lee has estimated the metallicity of the galaxy to be $[{\rm Fe/H}]=-2.10\pm 0.22$ and the distance, $0.81\pm 0.08$ Mpc from the width and position of the red giant branch (RGB) and the tip of the RGB (TRGB). In turn, Tikhonov \& Makarova have estimated a distance of $0.6\pm 0.1$ Mpc using the TRGB and the magnitude of the brightest stars in the galaxy. Mould also uses the TRGB to estimate a distance of $0.81\pm 0.15$. Finally, Karachentseva et al. have obtained a total apparent 
magnitude of $B_T=16.18$ and an axial ratio of 0.67.

Much effort has been devoted to the determination of the properties of LGS~3. 
The distance estimates indicate that it might be a satellite of M33. But the 
main interest of this object comes from the fact that, as mentioned above, it 
may be a transition object between a dIr galaxy and a low-luminosity dE (see 
Gallagher \& Wyse \markcite{gawy} 1994), having properties typical of both. If dwarf galaxies evolve from star forming to passive objects, LGS~3, together with other systems, like Phoenix (van de Rydt et al.~\markcite{vdrydt}, 1991), could be in a transition stage. LGS~3 shows a CM diagram similar to those of dE galaxies although with a small number of blue stars. It also has a small but 
significant amount of gas. Apparently, it is not currently forming stars or, at 
least, its present star formation rate (SFR) is very low, since the galaxy shows 
no traces of ionized gas. Its role as a possible missing link between the two 
subtypes of dwarf galaxies make a deep understanding of the 
properties of LGS~3 very important. Knowledge of its star formation history (SFH) will certainly help to reach that understanding. In this paper, we 
discuss what the SFH of LGS~3 may have been, using new ground-based photometry 
of its stars. 

In \S2 the observations are presented. Section~3 discusses the CM diagrams. 
In \S4 the distance is derived. Section~5 deals with the derivation of the 
SFH and \S6 is devoted to the discussion of the resulting SFH. In \S7 the 
chemical scenarios that would be compatible with the found SFH are investigated.
In \S8 integrated properties of the galaxy are discussed. In \S9 the main conclusions of the work are summarized.

\section{Observations and data reduction}

\noindent LGS~3 was observed in $V$, $R$ and $I$ (Johnson-Cousins) in 1993 October, at the 
Cassegrain focus of the Nordic Optical Telescope (NOT), at Roque de los 
Muchachos Observatory 
on the island of La Palma. The reader is referred to the paper by Aparicio \& Gallart 
\markcite{peg2} 
(1995) where details about this observing run are given. We give here a short 
description of the 
observations and data reduction. A thick, blue coated, Thomson CCD of 
$1024\times 1024$ pixels 
was used. It provided a pixel size of 0.14$''$ and a total field of $143\times 
143 ($''$)^2$. This 
field covered most of the galaxy optical body, although not completely. Total 
integration times 
were 6000 s in $V$, 3200 s in $R$, and 4600 s in $I$. Several 
frames were taken 
in each band, moving the telescope slightly (about 3 to 5 arcsec) between each 
two exposures, in 
order to reduce systematic pixel-to-pixel and position-dependent effects on the 
stellar 
magnitudes. The journal of observations is given in Table~1. The UT for 
the beginning of each exposure are given in column~1 all of them corresponding to 
October 17, 1993; column~2 lists the target (LGS~3 or 
a companion field); column~3 gives the filter; column~4 lists the integration 
times; and the 
full width at half maximum (FWHM) of the stellar profiles is given in column~5. 

The CCDRED package of IRAF was used to pre-process the data, using bias, dark 
and flatfield frames obtained during the observing nights. The effect of the 
dispersed light 
produced by the LED used to recognize the position of the filter wheel was also 
satisfactorily 
corrected (see Aparicio \& Gallart \markcite{noticias}1993 and 1995). DAOPHOT~II 
and ALLSTAR 
(Stetson \markcite{dao}1993) was used to derive the magnitudes of the stars, 
through multiple 
fitting of a point spread function (PSF) to the stellar profiles. An empirical 
PSF, varying 
with position on the frame, was obtained for each image. A Moffat analytical 
function and a 
table of residuals were computed, both obtained from a large number of 
well-shaped stars in each frame. The residuals of the fit, $\sigma$, and the CHI 
and SHARP parameters of ALLSTAR were used to 
select the stars with {\it good} photometry. Only stars with $\sigma <0.15$, 
CHI$\,<1.2$ and $-1.2<\,$SHARP$\,<1.2$ in all the 
frames ($V$, $R$ and $I$) were retained. All in all, 736 stars were kept, 
measured in at least 
two colors. Figure~\ref{sigdao} shows the plot of $\sigma$ values given by 
ALLSTAR for the 
retained stars. These plots are interesting because the $\sigma$ values are 
representative of the 
internal errors of the photometry as a function of magnitude: following Stetson 
\& Harris 
\markcite{stharr} (1988), the internal errors, defined as the $\sigma$ values 
obtained from 
frame-to-frame agreement, would not differ by more than about 20\% from the 
$\sigma$ values 
provided by ALLSTAR.

\placefigure{sigdao}

Aperture corrections were obtained using the image's FWHM and the relation given 
by Aparicio 
\& Gallart (1995), which are valid for the observing run under discussion:
$$m_{\rm psf}-m_{\rm ap}=-0.346+1.336{\rm FWHM},$$
\noindent where $m_{\rm psf}$ are magnitudes from PSF fitting and $m_{\rm ap}$ are 
aperture magnitudes. 
The $m_{\rm ap}$ were corrected for atmospheric extinction and then transformed into 
the 
Johnson-Cousins standard system, using the relations given in Aparicio \& 
Gallart (1995):
$$(V-v)=23.707-0.010(V-R)$$
$$(R-r)=24.344-0.021(V-R)$$
$$(I-i)=23.746+0.010(V-I)$$
\noindent where capital letters refer to Johnson-Cousins magnitudes and lower-case 
letters stand 
for  $m_{\rm ap}$ magnitudes above the atmosphere. These transformation equations, as 
well as the 
atmospheric extinctions in each band  and each night, were obtained from 
twenty-three 
measurements in each band of a total of 18 stars from the list of Landolt 
\markcite{landolt} 
(1992). The photometric conditions were particularly good, producing very small 
zero-point 
errors: 0.004 for $V$, 0.005 for $R$ and 0.009 for $I$. The main zero-point 
error source is 
therefore the $m_{\rm psf}-m_{\rm ap}$ correction. Taking all the error sources into 
account, we 
may estimate that our photometrical calibration is affected by total zero-point 
errors of about 
0.01 to 0.02 in the three bands. The full photometric list will be made 
available by the authors on request. Figure~\ref{imagen} shows the $I$ image.

\placefigure{imagen}

We have compared our photometry with the $VRI$ photometry of Lee (1995) and the $VI$ photometry of Mould (1997). No color terms are apparent but the following systematic zero points have been found:
$$V_{\rm Lee}-V_{\rm here}=-0.09\pm 0.01$$
$$R_{\rm Lee}-R_{\rm here}=+0.03\pm 0.02$$
$$I_{\rm Lee}-I_{\rm here}=-0.02\pm 0.02$$
$$V_{\rm Mould}-V_{\rm here}=+0.17\pm 0.01$$
$$I_{\rm Mould}-I_{\rm here}=+0.24\pm 0.01$$
About 25 stars with $V<22.5$ have been used in each case. Standard errors are quoted.

\section{The color-magnitude diagram}

Figures~\ref{obsvi} and \ref{obsvr} show the $I$ vs. $(V-I)$ and $V$ vs. $(V-R)$
CM diagrams of LGS~3. Contamination by foreground stars is very small for LGS~3, 
but may produce non-negligible effects on the star counts of the brightest, less 
populated parts of the CM diagram. To check this point, two additional frames of 
a field adjacent to LGS~3 were taken in $R$ and $I$ during the same observing 
run. Exposing times were 600 and 700 s, respectively (see Table~1). Only eleven 
stars have been resolved in the field. Colors and magnitudes of ten of these 
stars are smoothly distributed in the range $17<I<22.0$, $0.0<(R-I)<0.9$. The 
remaining star is a redder, fainter one and its photometry might be affected by 
a large random effect. This information will be used later on to correct the 
star counts in the CM diagrams. Since the integration times are short in the 
field frames, completeness might be significantly smaller than 1 at $I\simeq 
22.0$. But the number of foreground stars is so small that this cannot introduce 
any appreciable effect on the star counts in the lower part of the LGS~3 CM 
diagram.

\placefigure{obsvi}
\placefigure{obsvr}

The main feature of the LGS~3 CM diagrams is the clump that extends from 
$(V-R)\simeq 0.2$, $(V-I)\simeq 0.7$ at the faint magnitude limit, to 
$[(V-R),V]\simeq [0.8,22.0]$ and $[(V-I),I]\simeq [1.5,20.5]$. This feature is 
what we have been calling the {\it red-tangle} in previous papers (see Aparicio 
\& Gallart \markcite{peg1}1994). In principle, it is composed of the RGB and the 
asymptotic giant branch (AGB) locus of old and intermediate-age stars and by 
intermediate-age blue loops. The fact that the red-tangle is so narrow 
necessarily implies a small range of metallicities of the stars of LGS~3, although 
their ages may be in a large interval. We will discuss this in detail below in 
Sections 5 and 6.

For moderately high metallicities, of the order of a few thousands in $Z$, the 
AGB stars produce what we  call the {\it red-tail} (Aparicio \& Gallart 
1994). Its redward extension may be an indication of the metallicity of the 
system. Again LGS~3 has too few stars for the  redward 
extension to be clearly determined. The very bright stars at $I\simeq 18.5$, $(V-I)\simeq 2.6-2.8$ could 
be AGBs but they may also be foreground stars and AGBs would extend just up to 
$(V-I)\simeq 2.0$.

Young stars would be found in the main sequence (MS), the blue-loops or the red 
supergiant (RSG) phases. The few stars bluer than $(V-I)\simeq 0.6$ may be in 
one of the first two evolutionary phases (our companion field has some 2 stars 
in this area). The star at $I\simeq 18.0$, $(V-I)\simeq 1.2$ could be a RSG but, 
it could also be a foreground star.

\section{The distance}

The TRGB is the point of the CM diagram at which low-mass stars undergo the 
He-flash. The bolometric luminosity of this point is almost constant for 
different ages and moderate metallicities and therefore, it is an important distance 
estimator (Lee, Freedman \& Madore \markcite{distip} 1993). It is most valuable in 
galaxies like LGS~3, which have a low probability of having even a single Cepheid 
star. Unfortunately, the upper part of the RGB locus of LGS~3 is split into 
several small clumps in such a way that it is quite difficult to decide which of 
them actually marks the TRGB. Application of a Sobel filter $[-1,0,+1]$ to the luminosity function, after adequate smoothing, results in two clearly separated peaks close to the upper part of the RGB, at magnitudes $I_{\rm TRGB}=20.52\pm 0.08$ and $I_{\rm TRGB}=21.04\pm 0.05$. The first value seems more stable and corresponds to the upper cut-off of the RGB (see Fig.\ref{obsvi}). Errors are simple estimates of the wideness of the peaks at 1/3 of their maxima. Lee (1995) found the TRGB of LGS~3 at $I=20.4$, and Mould (1997) found it at $I=20.5\pm 0.4$. These values are close to our first possibility, although some traces of the discontinuity at $I\simeq 21.0$ can also be seen in Lee's and Mould's CM diagrams. Since the signal-to-noise ratio of our photometry at $I\sim 21$ is quite good and we have no strong criterion to decide, we will use both values for the 
magnitude of the TRGB ($I=20.52$ and $I=21.04$) and analyze the resulting 
properties of the galaxy in both cases. The stars over the TRGB are mainly AGB 
stars. This means that the use of either value for the magnitude of the 
TRGB can in principle significantly affect the results for the SFH, since many 
more AGBs would be present if $I_{\rm TRGB}=21.04$. 

To derive the distance modulus from the TRGB, the foreground extinction is 
needed. It can be estimated from B\"urstein \& Heiles \markcite{burstein} (1984) 
to be $A_B=0.16$, $A_V=0.12$, $A_R=0.09$, $A_I=0.07$. Using these values, the 
two possible de-reddened positions of the TRGB are 
$[(V-I)_0,I_0]_{\rm TRGB}=[1.33,20.45]$ and $[(V-I)_0,I_0]_{\rm TRGB}=[1.25,20.97]$. The 
distance modulus can now be calculated from the bolometric magnitude of the TRGB 
and an estimate of the metallicity. 
The latter can be obtained from the value of $(V-I)_0$ at $M_{I_0}=-3.5$ 
($\sim0.5$ magnitudes 
below the TRGB) using the relation given by Lee et al. (1993). The bolometric 
magnitude of the 
TRGB is then obtained as $M_{\rm bol}=-0.19[{\rm Fe/H}]-3.81$, which is transformed to the magnitude in the $I$ band using $M_I=M_{\rm bol}-{\rm BC}_I$, where ${\rm BC}_I$ is the corresponding bolometric 
correction derived from ${\rm BC}_I=0.881-0.243(V-I)_{\rm TRGB}$ (Da Costa \& Armandroff 
\markcite{dcar} 
1990). The resulting values of the distance modulus for each assumption of the 
TRGB are 
$(m-M)_0=24.43$ and $(m-M)_0=24.91$. The main error source is the uncertainty in 
$I_{\rm TRGB}$. 
Combined with the uncertainty in the extinction and in the calibration of the 
photometry, an 
estimate for the error of each of the former distance moduli is $\pm 0.2$ 
magnitudes. Taking this into account, the two values for the distance are 
$d=0.77\pm 0.07$ Mpc and $d=0.96\pm 0.07$ Mpc. In 
summary, we can only say that the distance to LGS~3 is compressed between 0.7 
and 1.0 Mpc. We 
will see later, from the results of the analysis of the SFH, that value 
$d=0.77\pm 0.07$ Mpc seems 
more likely. These intervals can be compared 
with the values previously obtained by Mould (1997): $0.81\pm 0.15$ Mpc, and by Lee (1995):  $0.81\pm 0.08$ Mpc, both from the TRGB and using the same method as here; by Tikhonov \& Makarova (1996) from the TRGB and luminosities of the brightest stars: $0.6\pm 0.1$ Mpc; and by Christian \& Tully \markcite{chris} (1983) from the brightest RSG: $0.7-1.2$ Mpc. 

\section{Derivation of the star formation history}

It may be said at first glance that the stellar population of LGS~3 may be  
composed by a number of old and intermediate-age stars, populating the  
red-tangle structure, plus a small amount of young stars, populating the bluest 
part of the CM diagram. The absence of a well developed red-tail can be 
interpreted as either the stars' metallicity always being very low or as the small 
number of stars resulting in a very low probability of having stars in the AGB 
phase. The amount of information in the LGS~3 CM diagram is less than in the 
cases of Pegasus (Aparicio \& Gallart 1995; Aparicio et al. 1997) 
or NGC~6822 (Gallart et al. 1996a,b,c) but several things can still be inferred 
about its SFH from the analysis that follows.

For the purposes of this paper, we consider the SFH to be a function of time 
accounting for all the characteristics which determine the formation of stars, 
mainly the star formation rate (SFR), the chemical enrichment law (CEL), and the 
initial mass function (IMF). Here we assume  the IMF to be fixed. We use the 
Kroupa, Tout \& Gilmore~\markcite{kroupa} (1993) function. Hence the SFR, which we will 
designate by $\psi(t)$ and the CEL, which we will designate $Z(t)$, both 
depending on  time, are the functions we will use to characterize the SFH. 

We have studied the SFH of LGS~3 via the comparison of its CM diagrams with 
model CM diagrams computed in a similar way as described in Gallart et 
al.~(1996b,c) and Aparicio et al.~(1997). Model CM diagrams are synthetic CM 
diagrams in which observational effects have been simulated. For this, the 
procedure explained in detail in Aparicio \& Gallart (1995) and Gallart et 
al.~(1996b) has been followed. In the case of LGS~3, 8000 artificial stars have 
been used to determine the observational effects. The $V$, $R$ and $I$ 
magnitudes of each star were simulated and they were added to the $V$, $R$ and $I$ images in groups of 40 stars at a time. DAOPHOT~II and ALLSTAR were run for each of the the 600 resulting frames to recover the magnitudes of the artificial stars. 
These recovered magnitudes, together with their input values are the information 
necessary to build up the {\it crowding trial table} (see Aparicio \& Gallart 
1995) used to simulate the observational effects in the model CM diagrams. Observational effects are, in particular, incompleteness effects, systematic shifts in color and magnitude as well as  internal and external errors, all them functions of stellar colors and magnitudes.

To analyze the SFH of LGS~3 with model CM diagrams, we have followed a different 
approach from that used for NGC~6822 (Gallart et al.~1996b,c) and Pegasus 
(Aparicio et al.~1996). We have first 
generated a small number of model CM diagrams produced by $\psi(t)$ functions 
defined as
$$\psi(t)=const. \hskip 10mm t_1\leq t<t_2$$
$$\psi(t)=0 \hskip 15mm t<t_1; t\geq t_2$$
\noindent and with metallicities randomly distributed in a given interval 
$Z_1\leq Z(t)<Z_2$. 
$t_1$, $t_2$, $Z_1$ and $Z_2$ are the inputs changing from model to model. In 
other words, these 
CM diagrams contain stars with ages and metallicities distributed in short 
intervals.  We will term these as {\it partial models} and denote their $\psi(t)$ functions by $\psi_i$. They will later be 
combined to produce what we will term 
{\it global models}, used to search the best representation of the observed CM 
diagram. We have 
first chosen the age intervals ($t_1$ and $t_2$) for each 
partial model. They are from 15 to 12 Gyr, 12--9 Gyr, 9--6 Gyr, 6--3 Gyr, 3--1 
Gyr and 1--0 Gyr 
(in practice the star formation is stopped at $t=0.01$ Gyr which is the minimum 
age available in 
the stellar evolution library used, see Bertelli et al.~\markcite{bert}1994). 

The color index 
and width of the  red-tangle is related firstly to the mean metallicity 
and the 
metallicity dispersion of the stars and secondly with the age distribution. We 
have used this 
information to determine what the metallicity range ($Z_1$ and $Z_2$) 
 should be that, given the 
age interval of each partial model, adequately reproduces both the position and 
width of the 
observed red-tangle. This has been done for all the age intervals except for the 
1--0 Gyr one, for which most of the evolved stars are not in the red-tangle. The 
metallicity for this interval has been obtained extrapolating the metallicity 
law derived using the remaining intervals. 

This approach to choose the metallicity 
law warrants a few comments. First, the reason we have adopted it is 
that we have no alternative information about $Z(t)$, lacking in 
particular measurements from HII regions or estimates from the structure of an 
AGB red-tail. Note furthermore that, in general, we use information of two kinds 
for studying the SFH: (i) position and overall shape of main structures in the 
CM diagram and (ii) distribution of stars inside and outside those structures 
(see Gallart et al. 1996b and Aparicio et al. 1997). Here, we are using 
information of the first kind to choose a valid $Z(t)$ law as the starting 
point. We no longer care about it in the rest of the process, in which we 
will use information of the second kind to put limits on $\psi(t)$ (see below). 
This strategy has the advantage of greatly reducing the number of partial models 
to be calculated, eliminating from the beginning all models which would 
produce a red-tangle with different width or in a different place from 
the observed one. It can be argued, however, that assuming that 
the entire width of the red-tangle is due to the metallicity dispersion will 
result in obtaining a SFH with a smaller dispersion in age. 
This is obviously true, but we prefer to adopt this position rather than the 
opposite, that would be to assume that stars at a given age do not
have any metallicity dispersion, which would favor 
a larger age dispersion. If LGS~3 were an intermediate case between a dIr and a 
dE, we might expect it to show an age dispersion smaller than that of a dIr. By assuming that the entire width of the red tangle is due to the metallicity dispersion, we are favoring to find a small age dispersion. If, in spite of this fact, we 
still find a likely large age dispersion, we are in a better position to believe 
that such is really the case.   

\placefigure{zage}

In summary, with the adopted procedure, we reduce the number of partial model CM 
diagrams to be computed to only 6 for each of the two possible distances. 
Figure~\ref{zage} shows the $Z(t)$ laws corresponding to these distances. $Z(t)$ 
functions significantly different from these would produce model CM diagrams that do not 
match the observed one. Note that the only thing assumed at this level is 
that, if LGS~3 has stars in a given age interval, their metallicities should be 
in the corresponding range shown in Fig.~\ref{zage}.

\placefigure{8reg}

The comparison between model and observed CM diagrams intended to put limits on 
$\psi(t)$ is done through star counts in the eight regions of the CM diagrams 
defined in Fig.~\ref{8reg}. These regions have been chosen to sample specific 
stellar evolutionary phases. Region~1 includes MS and blue-loop stars. Region~2 
would be populated by RSGs and by bright AGBs. Regions~3 and 4 correspond to the 
part of the diagram where intermediate-age and old AGBs are expected to appear 
if the metallicity is not very low. Region~5 is defined just over the TRGB and would 
contain low-metallicity AGB stars. Regions~6, 7 and 8 sample the red-tangle. 
Figure~\ref{8reg} corresponds to $d=0.77$ Mpc. Regions are defined at the same positions in $[(V-I)_0,M_{\rm I}]$ for $d=0.96$ Mpc. In this case, region~8 is not considered because its 
lower boundary is too close to the limit of the photometry.

\placefigure{nstar}

The numbers of stars in each of the former regions of the observed $I$ vs. 
$(V-I)$ CM diagram are shown in Fig.~\ref{nstar} (solid lines) for both 
distances. These numbers have been corrected from foreground contamination using 
the information given in Table~2. This table contains  the 
number identifying each region in the first column. The second column gives the adopted numbers of foreground stars contaminating each region. They have been determined from the photometry performed in the comparison field (see \S~2). The numbers of foreground stars are very small, but can be relevant in the less populated 
regions. A single star appears in the foreground comparison field around the 
lower limit used to define region~3. This star is outside the region, but could 
be inside if the error intervals for the distance and the photometry are 
considered. For this reason, the value 0.5 has been adopted for the foreground 
stars in region~3.  

Global models are obtained as a combination of partial models. They have arbitrary 
$\psi(t)$ functions produced by choosing the adequate fraction of stars of each 
partial model. Global models are those to be compared with the observed CM 
diagram to search for the SFH. But the global models do not need to be 
explicitly computed. Once the amount of stars with which each partial model 
populates 
each of the regions defined in the CM diagram is known, the amount of stars that 
populate the 
same regions for a given global model can be directly found as a linear 
combination of the form
$$N_{g,j}=\sum_{i=1}^ma_ik_iN_{i,j},$$
\noindent where $N_{g,j}$ is the number of stars in region $j$ for the global 
model; $N_{i,j}$ is the same number corresponding to partial model $i$; $k_i$ 
normalize each partial model to the same constant $\psi(t)$; $a_i$ are the 
coefficients defining the global model and $m$ is the number of partial models 
used (six in our case). In other words, the $N_{g,j}$ values are all 
we need for comparing the global models with observations. An arbitrarily large 
number of global models can be made by just changing $a_i$. 

In practice, we have done the following. First, we have chosen for $a_i$ the values 
0, 0.5, 1, 1.5, 2, 3, 5, 8, 12 and 18 and we have computed all the $N_{g,j}$  
corresponding to the one million possible global models resulting from giving to 
$a_i$ all the possible variations of the former values. The $N_{g,j}$ are then 
normalized to reproduce the total number of stars in the observed CM diagrams 
inside the regions used:
$$l\sum_{j=1}^rN_{g,j}=\sum_{j=1}^rN_{o,j}.$$
\noindent where $r$ is the total number of regions (seven or eight in our case) 
and $N_{o,j}$ is the number of stars in region $j$ in the observed diagram. The absolute scale of $\psi(t)$ is set by this normalization; the particular values given to $a_i$ mean only relative weights of the SFR at each epoch. 

The SFR corresponding to each global model is then
$$\psi(t)=l\sum_{i=1}^ma_ik_i\psi_i$$
\noindent where $\psi_i$ corresponds to partial model $i$. 

An error can be associated to each of $N_{g,j}$ and $N_{o,j}$ assuming Poissonian 
distributions of data. 
For the observed values the errors are simply
$$\sigma_{o,j}=N_{o,j}^{1/2}$$
\noindent Errors of models have to include the different normalizations. 
First, the errors of 
partial models are
$$\sigma_{i,j}=k_iN_{i,j}^{1/2}.$$
\noindent Errors of global models are
$$\sigma_{g,j}=l(\sum_{i=1}^ma_i^2\sigma_{i,j}^2)^{1/2}.$$

Then, global models producing stars counts in each of the $r$ regions of the two 
CM diagrams [$I$ vs. $(V-I)$ and $V$ vs. $(V-R)$] differing from the star 
counts of the observed CM diagrams by less than a given number of $\sigma$ are 
selected; i.e., the models verifying
$$\delta_j\leq n\sigma_j \hskip 1cm \forall j,$$
\noindent where $\delta_j=$abs$(N_{g,j}-N_{o,j})$, 
$\sigma_j=(\sigma_{g,j}^2+\sigma_{o,j}^2)^{1/2}$ and $n$ is the allowed error 
interval, expressed in $\sigma$. 

\section{The Resulting Star Formation History}

We have performed the former procedure for the two choices of the distance. 
There is no global model compatible with observations at the $1\sigma$ level; i.e., 
there is no global model which reproduces the number of stars in each of the $r$ 
(seven or eight) regions of both CM diagrams to better than $1\sigma$. 2.5\% of 
global models are compatible with observations at the $1.33\sigma$ level for 
$d=0.77$ Mpc but models up to $2\sigma$ have to be accepted to have a similar 
amount in the case of $d=0.96$ Mpc. No models are compatible with 
observations at $1.33\sigma$ for this distance. 

The SFH of any global model selected in this way would produce a CM diagram 
compatible with observations at the quoted level. A general representation can be provided by the average of all the selected models. Figure~\ref{best_ca} shows the average 
$\psi(t)$ functions of models accepted at $1.33\sigma$ for distance $d=0.77$ Mpc 
and $2\sigma$ for distance $d=0.96$. Error bars show the dispersion of $\psi(t)$ 
for each age interval. In Figure~\ref{nstar}, 
the number of stars in the eight regions used in the observed $I$ vs. $(V-I)$ diagrams are compared with those produced by the averaged models. Similar representations are obtained for the $V$ vs. $(V-R)$ diagrams. The fact that results are worse for $d=0.96$ Mpc might be an indication that the right distance is 0.77 Mpc and that, therefore, the discontinuity in the RGB at $I\sim 21.00$ is a random effect. This distance agrees with the values found by 
Mould (1997) and Lee (1995), but it must be checked by another independent procedure.

\placefigure{best_ca}
\placefigure{posibles}

Figure~\ref{best_ca} shows that an almost constant (for $d=0.77$ Mpc) or decreasing (for $d=0.96$ Mpc) $\psi(t)$ can reproduce the observed CM diagrams. But the large error bars indicate that, for $t>1$ Gyr, $\psi(t)$ can vary within very wide intervals. However,  $\psi(t)$ cannot move inside the error intervals in an arbitrary way.  Figure~\ref{posibles} gives complementary information for the case of $d=0.77$ Mpc. To plot panel a), the accepted global models have been divided into three sets, according to their $\psi(t)$ for the time interval 15 to 12 Gyr: those having $\psi(t)$ larger than $0.5\sigma$ over the average value; those having $\psi(t)$ closer than $0.5\sigma$ to the average and those having $\psi(t)$ smaller than $0.5\sigma$ below the average value. In other words, they have divided into models with high, intermediate and low $\psi(t)$ during the 3 first Gyrs. The same criterion has been applied to plot the rest of the panels, but using the value of $\psi(t)$ in different time intervals: panel b) from 12 to 9 Gyr; panel c) 9--6 Gyr; panel d) 6--3 Gyr; panel e) 3--1 Gyr. No division has been done for the interval from 1 to 0 Gyr because the value of $\psi(t)$ is very stable for this last period. The integrated $\psi(t)$ must account for the total number of stars in the observed CM diagram. Therefore, if $\psi(t)$ is higher in a given time interval, it must be compensated for by lower $\psi(t)$ in other time intervals. This can be seen in Fig.~\ref{posibles}. The interesting point is that the effects of choosing a higher or lower $\psi(t)$ in a given time interval are mostly compensated for by adjustments in $\psi(t)$ in the adjacent intervals. This indicates that our actual time resolution is poorer than the 3 Gyr interval we have used; i.e., we can reproduce the observations for example, having many stars older than 12 Gyr and a few between 9 and 12 Gyr or viceversa. Nevertheless, coupling seems stronger between intervals from 15 to 9 Gyr and between intervals from 9 to 1 Gyr. For this reason we have averaged $\psi(t)$ in these wider intervals. The resulting function is much better constrained. This can be seen in Fig.~\ref{best_ca2}, which shows that:

\placefigure{best_ca2}

\begin{itemize}
\item For $d=0.77$ Mpc, $\psi(t)$ must not have been very different from constant from 15 to 1 Gyr ago and it has decreased by a factor of $\sim 3$ in the last 1 Gyr. The estimate by Mould (1997) of a decreasing of $\psi(t)$ by a factor 10 seems too large in the light of our models.
\item For $d=0.96$ Mpc, $\psi(t)$ decreases by factors of $\sim 2$ between the periods running from 15 to 9 Gyr, from 9 to 1 Gyr and from 1 to 0 Gyr. Defining $\tau=15-t$, $\psi(\tau)$ can be fitted by an exponential law of the form
$$\psi(\tau)=\psi_0e^{-\tau/\beta}$$
\noindent with $\psi_0=1.5\times 10^{-4}$ M$_\odot$yr$^{-1}$ and $\beta=7.1$ Gyr.
\end{itemize}

We have seen that time resolution in the interval from 15 to 9 Gyr is poor, so that we hardly could say anything about the precise time of the beginning of the star formation in LGS~3. The fact that 100\% of the accepted models have star formation in that interval is a strong indication of an early epoch for the beginning of the star formation, reinforced by the fact that 97\% of the 
accepted models have star formation from 15 Gyr ago. Note for comparison that 90\% of the computed models have star formation since 15 Gyr ago. These figures are valid for both distances. 

A similar question can be raised about $\psi(t)$ at present date. The fact that LGS~3 has no HII regions could indicate that $\psi(0)=0$. But the values found for the last 1 Gyr ($(1.5\pm 0.1)\times 10^{-5}$ M$_\odot$ yr$^{-1}$ for $d=0.77$ Mpc and $(1.9\pm 0.3)\times 10^{-5}$ M$_\odot$ yr$^{-1}$ for $d=0.96$ Mpc) combined with the IMF by Kroupa et al.~(1993), results in a probability of LGS~3 having 
a star massive enough to produce an HII region of about 0.2, compatible with observations.

\placefigure{popbox}
\placefigure{sinfin}

Figure~\ref{popbox} shows the population box (see Hodge 1989) for LGS~3 plotted 
using the results of distance 0.77 Mpc. The figure simultaneously shows 
$\psi(t)$ and $Z(t)$ and is a global representation of the SFH of the galaxy.
Figure~\ref{sinfin} shows the model CM diagrams produced by the SFH plotted in Fig.~\ref{popbox}. These diagrams can be considered the ones better reproducing the observed diagrams shown in Figs.~\ref{obsvi} and \ref{obsvr}.

\placefigure{hst}

Figure~\ref{hst} shows a model CM diagram produced by the SFH plotted in Fig.~\ref{popbox} also, but simulating the observational effects expected if the limiting magnitude were $\sim 3$ magnitudes deeper than in our data. This is what would be obtained from deep HST observations, if our conclusions are correct. Note that the well developed HB is produced by the presence of a significant amount of very old, low metallicity stars in the adopted model. Eventual differences in the observed HB would imply a different SFH for the first $\sim 1$ Gyr. It is in fact the morphology of the HB which would provide details on the SFH for that very early period. Similarly, the distribution of MS stars will provide details about the SFH in the last $\sim 1$ Gyr.

\section{The Chemical Enrichment Scenario}

We have calculated $\psi(t)$ and $Z(t)$ laws that reproduce the CM diagram of LGS~3. Now we will investigate what chemical evolutionary scenarios make compatible the $\psi(t)$ and $Z(t)$ found, i.e. under what overall conditions $\psi(t)$ as shown in Fig.~\ref{best_ca2} produce $Z(t)$ as shown in Fig.~\ref{zage}. For each distance, we have computed the $Z(t)$ that would be produced by our $\psi(t)$ for a simple closed box scenario and for families of infall and outflow scenarios. Following Peimbert, Col\'\i n, \& Sarmiento\markcite{peimbert} (1994, and references therein) we assume the infall rate as given by $f_{\rm I}=\alpha(1-R)\psi$ and the outflow of well mixed material by $f_{\rm O}=\lambda(1-R)\psi$, where $R$ is the returned fraction (Tinsley~\markcite{tinsley} 1980) and $\alpha$ and $\lambda$ are parameters. Outflow of non mixed $Z$-rich material has been also considered and is characterized by the parameter $\gamma$, defined as the fraction of 
freshly made metals that are ejected to the intergalactic medium without mixing with the interstellar gas. A returned fraction $R=0.2$, accordingly to our stellar evolutionary models, and a gas fraction $\mu=0.4$ have been used (see \S8 and Table~3). Figure~\ref{cel} shows a number of the chemical enrichment laws we have computed in this way, overimposed to the $Z(t)$ laws found for LGS~3 shown in Fig.~\ref{zage}.

\placefigure{cel}

In the case of $d=0.77$ Mpc, a closed box produces a $Z(t)$ law not very different to that we have found for LGS~3, but a better fit is obtained with a moderate outflow of well mixed material given by $\lambda=3$. In contrast, large rates of infall or outflow would produce $Z(t)$ increasing too fast and too slowly, respectively, in the beginning of the galaxy history. The effective yield obtained for the $\lambda=3$ scenario is $y=0.002$ (it is $y=0.0011$ in the closed box case), 7 times smaller than the theoretical value $y=0.014$ found by Peimbert et al. (1994) using models by Maeder~\markcite{maeder} (1992) for a system with $Z=0.001$. Effective and theoretical yields can be reconciled if, in addition, outflow of $Z$-rich material at a rate $\gamma=0.85$ is allowed. Since the material escaping from the galaxy without mixing would mostly be that coming from SN explosions, this would modify the rates of chemical elements present in the interstellar medium and successive generations of stars. In particular, following the results by Peimbert et al. (1994, and references therein) the ratio $\Delta Y/\Delta Z({\rm O})$ would be 4 times larger than if no outflow of $Z$-rich material would exist. This value is not far from the factor 2.4 (which corresponds to $\gamma=0.73$) found by Peimbert et al. (1994) for a sample of irregular and blue compact galaxies (see also Pagel et al.~\markcite{pagel} 1992). 

In the case of $d=0.96$ Mpc, a large outflow of well mixed material at a rate  $\lambda\simeq 30$ is required to reconcile the $\psi(t)$ and $Z(t)$ functions of LGS~3 (see Fig.~\ref{cel}). The effective yield is now $y=0.009$. Additional outflow of $Z$-rich material at a rate $\gamma=0.36$ would reconcile this yield with the theoretical one.

\section{Integrated properties of LGS~3}

The results found for $\psi(t)$ can be used to calculate integrated properties. 
The most relevant ones are summarized in Table~3 for both distances. $\bar\psi$ 
is the average of $\psi(t)$ from 15 Gyr ago to date. We cover in our 
observations almost the full optical body of the galaxy, but not completely. In 
this sense, the actual value of $\bar\psi$ should be slightly 
larger than that given in the table. The value of $\bar\psi$ normalized to the 
observed area is also given ($\bar\psi/A$). $\bar\psi_{1{\rm Gyr}}$ is the average 
value of $\psi(t)$ for the last 1 Gyr and  is also given per unit area 
($\bar\psi_{1{\rm Gyr}}/A$). $Z(0)$ is the current metallicity. $M_\star$ is the mass locked into stars and stellar 
remnants obtained from integration of $\psi(t)$ and using the IMF of Kroupa et 
al. (1993). $M_{\rm gas}$ and $M_{\rm tot}$ have been calculated using the results by Lo et al.~(1993) and assuming the gas mass to be 4/3 of the HI mass. To compute the $B$ luminosity, $L_B$, the apparent magnitude $B_T=16.18$ by Karachentseva et al. (1997) has been used. For the galactic extinction we have used $A_{Bg}=0.16$ 
(see \S~4) and we have computed the internal extinction as $A_{Bi}=0.75\log 
R$ (de Vaucouleurs et al. \markcite{dvau}~1991) where $R=1/0.67$ for LGS~3 
(Karachentseva et al. 1997). For the solar absolute magnitude, $M_{B\odot}=5.54$ 
has been used, after Durrant \markcite{durrant}~(1981). $\mu$ is the fraction of 
gas relative to the total mass intervening in the chemical evolution: 
$\mu=M_{\rm gas}/(M_\star+M_{\rm gas})$. As dark matter, we consider the mass that cannot be explained as stellar remnants or by extrapolation of the Kroupa et al. 
(1993) IMF to low masses. The fraction of dark matter is computed as 
$1-(M_\star +M_{\rm gas})/M_{\rm tot}$. $M_{\rm b,0}$ is the initial mass intervening in the chemical evolution. It is computed as the current value of $M_\star +M_{\rm gas}$ corrected of the outflow found in \S7. This outflow is $(\lambda+\gamma{R\over (1-R)}) M_\star$, where we have used $R=0.2$, according to ours models. In the case $d=0.77$ Mpc, $\lambda=3$ and $\gamma=0.85$. In the case $d=0.96$ Mpc, $\lambda=30$ and $\gamma=0.36$. Note that the value $M_{\rm b,0}=2.0\times 10^7$ M$_\odot$ obtained for $d=0.96$ Mpc is close to the dark matter mass ($2.1\times 10^7$ M$_\odot$), indicating that, if this mass is constant in the galaxy, its amount could have been about 50\% of the total initial mass of the galaxy.

The next two lines of Table~3 give the gas and total 
mass-luminosity relation. The last three lines contain the distances of LGS~3 to 
M33, to M31 and to the barycenter of the LG. To calculate this, we have used the 
distance from the Milky Way to M33 and M31 to be respectively 0.84 Mpc (Freedman, Wilson \& Madore \markcite{freed_m33}~1991) and 0.77 Mpc (Freedman, \& Madore 
\markcite{freed_m31}~1990) and the barycenter of the LG to be situated in the 
line connecting M31 and the Milky Way, at a distance of 0.45 Mpc from the Milky 
Way. This results from adopting, after Peebles \markcite{peebles}~(1989), a mass 
for the Milky Way 0.7 times that of M31 and neglecting the masses of any other 
galaxy in the LG.

Let us go back to the idea that LGS~3 might be a dIr-dE intermediate-type galaxy. 
As we have mentioned, indications for this are that the galaxy shows a small 
amount of gas but neither HII regions nor a well populated MS. However, we have 
seen that LGS~3 shows an $M_{\rm gas}/L_B$ ratio close to 1 and that its CM diagram 
is compatible with a roughly constant star formation starting 15 Gyr ago and with a 
present rate not much lower than its average for the past. Furthermore, LGS~3 
has enough gas to be forming stars for some 30 Gyr more from now at the present 
rate or for some 10 Gyr more from now at the averaged rate $\bar\psi$. All this 
would not be surprising for a dIr. The characteristic of LGS~3 which makes it 
different from other dIrs are its absolute parameters. $M_{\rm tot}$, $M_{\rm gas}$, 
$M_\star$, $L_B$ and $\bar\psi$ are one or two orders of magnitude smaller that 
those for typical dIr galaxies. As a result, we conclude that LGS~3 does not show any HII for reasons of simple probability and not necessarily because it is in a quiescent 
phase of its star formation. In summary, LGS~3 might well be a dIr just on the 
tail of the mass, luminosity and SFR distributions.

\section{Conclusions}

Photometry in the $V$, $R$ and $I$ Johnson-Cousins bands of 736 stars resolved 
in LGS~3 is presented. The galaxy shows CM diagrams where the main feature is a narrow 
red-tangle, but very few red-tail or blue stars are present. The distance of the galaxy is estimated using the luminosity of the TRGB. A splitting of the upper part of the RGB prevents a unique determination and forces the assumption of two possible values for the distance: $d=0.77\pm 0.07$ Mpc and $d=0.96\pm 0.07$ Mpc. From the analysis of the SFH, indications that $0.77\pm 0.07$ is the best estimate are obtained.

The SFH of the galaxy is investigated. We consider the SFH as the combination of 
two time-dependent functions: the SFR $\psi(t)$ and the CEL $Z(t)$, the IMF 
being assumed as a fixed input. The SFH has been studied through the comparison 
of the observed CM diagrams with model CM diagrams which result from different 
choices of $\psi(t)$ and $Z(t)$. The following approach has been used: first, the 
numbers of stars produced by a small number of simple-shaped star formation 
rates $\psi(t)$ (we call them $\psi_i$) are computed in several regions of the CM diagram. Each $\psi_i$ produces stars at a constant rate for a short interval of time, and in a given metallicity interval. The distribution of stars in the CM diagram 
produced by an arbitrary $\psi(t)$ is then computed as a linear combination of 
the results for the $\psi_i$ functions, as explained in \S5. We call 
each of these linear combinations a {\it global model}. In total, $10^6$ global 
models have been calculated. We have selected all those that reproduce the 
numbers of stars in previously defined regions of the observed CM diagrams 
better than $1.33\sigma$ for $d=0.77$ Mpc and $2\sigma$ for $d=0.96$ Mpc. 2.5\% of the models fulfill each condition. 

The $\psi(t)$ functions resulting from average of all the accepted models are 
represented in Figs.~\ref{best_ca2} and \ref{popbox}. The results can be 
summarized as follows:

\begin{itemize}

\item For both possible distances star formation beginning 15 Gyr or, at 
least, 12 Gyr ago is favored. Nothing can be said about short time bursts in 
the past, since the CM of LGS~3 does not give enough resolution for ages older 
than 1 Gyr. 
\item An almost constant function from 15 to 1 Gyr ago and stepping down by a factor $\sim 3$ since 1 Gyr ago (in the case of $d=0.77$ Mpc) or an exponentially decreasing function with $\beta=7.1$ Gyr extending from 15 Gyr ago to date (in the case of $d=0.96$ Mpc) are likely shapes for $\psi(t)$. 
\item The average value of $\psi(t)$ for the entire lifetime of the galaxy is 
$\bar\psi=(1.4\pm 0.1)\times 10^{-10}$ M$_\odot$ yr$^{-1}$ pc$^{-2}$ for both distances.
\item The current metallicity ranges from $Z\simeq 0.0007$ to $Z\simeq 0.002$ for $d=0.77$ Mpc and from $Z\simeq 0.0007$ to $Z\simeq 0.0015$ for $d=0.96$ Mpc 
\item Using $\bar\psi_{\rm 1Gyr}$ as a good representation of the present SFR, the 
probability of LGS~3 having an HII region is about 0.2, compatible with the 
observation of no HII regions in the galaxy.
\item The gas fraction of LGS~3 relative to the mass intervening in the 
chemical evolution is about 0.4. 
\item About 95\% of the mass of LGS~3 is dark matter, i.e., mass that is neither 
explained by stellar remnants nor by an extrapolation of the Kroupa et al. 
(1993) IMF to low-mass stars.
\item Outflow of well mixed material is required to make compatible the $\psi(t)$ and $Z(t)$ found. For the case of $d=0.77$ Mpc, a relatively moderate outflow given by $\lambda=3$ is enough, but $\lambda\simeq 30$ is required for the case $d=0.96$ Mpc. Effective yields are smaller than theoretical ones derived from stellar evolution models of Maeder (1992). Outflow of a fraction $\gamma=0.85$ of non mixed, freshly made metals, in the case $d=0.77$ Mpc, and of $\gamma=0.36$ in the case $d=0.96$ Mpc, would account for the differences between effective and theoretical yields.
\item The large outflow rate $\alpha\simeq 30$ required in the case $d=0.96$ Mpc would imply that most of the initial gas of the galaxy has been lost and that the initial dark matter fraction would have been as low as $\sim 50$\%.
\item The $M_{\rm gas}/L_B$ ratio is 0.74, similar to what is common in dIr 
galaxies. With the observed $M_{\rm gas}$, LGS~3 can continue to form stars for some 30 Gyr more at the present rate or for some 10 Gyr more at a rate equal to 
$\bar\psi$.
\item The distances of LGS~3 to M33 and M31 are respectively 0.18 and 0.25 Mpc if $d=0.77$ Mpc, or 0.22 and 0.34 Mpc if $d=0.96$ Mpc.

\end{itemize}

In summary, LGS~3 shows characteristics of typical dIrs (gas fraction, $M_{\rm gas}/L_B$ rate, behaviour of $\psi(t)$), the main differences being that its mass, luminosity and 
SFR (present and averaged) are one to two orders of magnitude too 
small. This makes the absence of HII regions a simple probabilistic effect. 
Taking into account all the properties of LGS~3, it may be considered a dIr just 
on the tail of the distributions of mass, luminosity and SFR, and not necessarily a dIr-dE transition object.

\acknowledgments

This work has been substantially improved after fruitful suggestions made by an anonymous referee. This work has been financially supported by the Instituto de Astrof\'\i sica de Canarias (grant P3/94) and by the Direcci\'on General de Investigaci\'on Cient\'\i fica y T\'ecnica of the Kingdom of Spain (grant PB94-0433).

\newpage
\figcaption[sigdao]{Residuals of the fitting of the PSF models vs. magnitude, as 
provided by ALLSTAR, for the stars included in the final photometric list of LGS~3.
\label{sigdao}}
\figcaption[imagen]{$I$ image of LGS~3. The total field is $143\times 143$ 
($''$)$^2$ and the integration time 4600 s. North is {\it up}, east is  {\it left}.
\label{imagen}}
{\figcaption[obsvi]{Observational $I$ vs. $(V-I)$ CM diagram of LGS~3.
\label{obsvi}}
\figcaption[obsvr]{Observational $V$ vs. $(V-R)$ CM diagram of LGS~3.
\label{obsvr}}
\figcaption[zage]{The adopted metallicity laws for each of the two possible 
distances. Upper panel shows the region of the metallicities 
allowed for the case of distance $d=0.77$ Mpc. Stars of a given age in the models have 
their metallicities compressed in this region. Lower panel shows the same for the case $d=0.96$ Mpc. Laws which significantly differ 
from these would produce model CM diagrams not matching the observed one.
\label{zage}}
\figcaption[8reg]{The 8 regions used for the determination of $\psi(t)$ 
superimposed on LGS~3's $I$ vs. $(V-I)$ and $V$ vs. $(V-R)$ CM diagrams 
corrected for reddening and for a distance modulus of $(m-M)_0=24.43$ ($d=0.77$ 
Mpc).
\label{8reg}}
\figcaption[nstar]{Number of stars in each region of the $I$ vs. $(V-I)$ CM 
diagram for the two distances considered. Full lines refer to the observed 
diagram. Dashed lines correspond to the averages of all the accepted models.
\label{nstar}}
\figcaption[best_ca]{Average of accepted $\psi(t)$ of LGS~3. Upper panel corresponds to $d=0.77$ Mpc and has been obtained averaging all 
models compatible with the corresponding observed CM diagrams at 1.33$\sigma$ 
level. Lower panel corresponds to $d=0.96$ Mpc and has been obtained 
by averaging all models compatible with the corresponding observed diagrams at 
the 2$\sigma$ level. 
\label{best_ca}} 
\figcaption[posibles]{Average of possible $\psi(t)$ functions considering several subsets according to the value of $\psi(t)$ (high, medium or low) at each of five time intervals. In each panel, dashed, full and dotted lines respectively correspond to averages of models having high, medium and low $\psi(t)$ at the time interval marked by stars (see text for details).
\label{posibles}}
\figcaption[best_ca2]{Likely SFR of LGS~3 as a function of time, obtained from the same data plotted in Fig.~\ref{best_ca} after rebinning time intervals. Upper panel corresponds to $d=0.77$ Mpc. Lower panel corresponds to $d=0.96$ Mpc.
\label{best_ca2}}
\figcaption[popbox]{The population box of LGS~3 represents the SFH of 
the galaxy combining the two key functions: $\psi(t)$ and $Z(t)$. Results for distance $d=0.77$ Mpc have been used because this seems the most likely distance for the galaxy.
\label{popbox}}
\figcaption[sinfin]{Model CM diagrams produced by the SFH shown in Fig.~\ref{popbox}. They are the ones better reproducing observations (Figs.~\ref{obsvi} and \ref{obsvr}).
\label{sinfin}}
\figcaption[hst]{Model CM diagram produced by the SFH shown in Fig.~\ref{popbox} simulating the observational effects expected if the limiting magnitude were $\sim 3$ magnitudes deeper than in this paper. This is what would be seen from deep HST observations if our conlusions for the SFH are correct.
\label{hst}}
\figcaption[cel]{Chemical enrichment laws for different scenarios produced by the $\psi(t)$ functions shown in Fig.~\ref{best_ca2} superimposed to the $Z(t)$ laws found for LGS~3. In both panels, dots mark the centers of the $Z(t)$-$t$ boxes. Full lines correspond to a closed box scenario; dotted lines to an infall scenario in which infall is balanced by star formation ($\alpha=1$) and dashed lines to an outflow scenario in which outflow is 30 times the mass locked into stars and stellar remnants ($\lambda=30$). In the upper panel, the dot-dashed line corresponds to an outflow scenario with $\lambda=3$.
\label{cel}}

\end{document}